# Fermionic analogue of black hole radiation with a super high Hawking temperature

Hang Liu,[1,5] Jia-Tao Sun,[1,2,5,*] Huaqing Huang,[3] Feng Liu,[3,4,†] and Sheng Meng[1,4,5,‡]

[1] *Beijing National Laboratory for Condensed Matter Physics and Institute of Physics, Chinese Academy of Sciences, Beijing 100190, People's Republic of China*

[2] *School of Information and Electronics, Beijing Institute of Technology, Beijing 100081, P. R. China*

[3] *Department of Materials Science and Engineering, University of Utah, Salt Lake City, Utah 84112, USA*

[4] *Collaborative Innovation Center of Quantum Matter, Beijing 100084, People's Republic of China*

[5] *University of Chinese Academy of Sciences, Beijing 100049, People's Republic of China*

**Abstract:** Measurement of gravitational Hawking radiation of black hole (BH) is prohibitive because of an extremely low Hawking temperature ($T_H$). Here we demonstrate a fermionic analog of BH with a super high $T_H \sim 3$ K, which is several orders of magnitude higher than previous works. We propose that Floquet-Dirac states, formed in a periodically laser driven two-dimensional black phosphorous thin film, can be designed with a spatial gradient to mimic the "gravity" felt by fermionic quasiparticles as that for a Schwarzschild BH (SBH). Quantum tunneling of electrons from a type-II Dirac cone (inside BH) to a type-I Dirac cone (outside) emits a SBH-like Hawking radiation spectrum.

Gravitational black hole (BH) is a curved spacetime absorbing everything inside its event horizon because of an extremely large mass/radius ratio, as characterized by overtilted light cones according to general relativity theory. This notion of nothing escapes from a BH is no longer valid if quantum effects are considered. According to Hawking [1-3], quantum fluctuations at the event horizon generate particles and antiparticles to propagate out of and into the BH, respectively. Consequently, a BH evaporates thermally like a black body, known as Hawking radiation, as shown in the upper panel of Fig. 1. The intensity of this quantum radiation is quantified by Hawking temperature ($T_H = \kappa/2\pi$ with the gravity $\kappa$ at the event horizon), which is however extremely low, ~$10^{-8}$ K for a BH with one solar mass [1]. This prohibitively weak intensity presents a significant challenge to directly observe Hawking radiation. Previously, artificial analogs have been proposed including sonic BHs [4, 5] in Bose-Einstein condensates [6-9], ion rings [10], and Fermi-degenerate liquids [11]. Unfortunately, $T_H$ for most artificial BHs is still very low, e.g. ~ $10^{-9}$ K in Bose-Einstein condensates [7]. Furthermore, the observation from some optical BH analogs remains controversial [12-21]. Therefore, new BH analogs inherent with quantum effect and a much higher $T_H$ are highly desirable.

Recently, type-II Weyl/Dirac fermions in solids have been proposed as a new platform to realize artificial BHs [22-27]. In the regions where type-II and type-I fermions are separated by a boundary with type-III fermions (i.e. the "event horizon"), two worlds inside and outside a BH are analogously formed (lower panel of Fig. 1). Quantum mechanics enables Hawking radiation of quasiparticles at the event horizon, which can simulate fermions radiation from a real BH [28-30]. Thanks to the fact that electrostatic interactions in solids are orders-of-magnitude stronger than gravitational forces, an unprecedented high $T_H$ is expected. Our previous work shows that laser-driven black phosphorus (BP) can host type-I, -II and -III fermions [25], manifesting a potential fermionic analogue of BH. In order to materialize this idea two requirements are necessary: (i) a spatial distribution of band structure to produce a "steep" analogous gravity field, and (ii) a working mechanism to induce fermionic Hawking radiation.

In this article, we theoretically propose a fermionic analogue of BH in two-dimensional

(2D) BP thin film under laser illumination, which is designed to be experimentally accessible. Combining first-principles and quantum tunneling calculations, a spatially inhomogeneous system with successively distributed type-II, -III and -I Dirac fermions is illustrated, which acts like an Schwarzschild BH (SBH) metric to induce electron emission from type-II to type-I region. An effective gravity field corresponding to a striking high temperature $T_H \sim 3$ K is achieved.

Dirac fermions are classified into different types, and successive transitions between them in 2D can be described by the Hamiltonian

$$H(\boldsymbol{k}) = c_x k_x \sigma_x + c_y k_y \sigma_y + v k_y \sigma_0 ,\qquad(1)$$

where $\sigma_x = \begin{pmatrix} 0 & 1 \\ 1 & 0 \end{pmatrix}$, $\sigma_y = \begin{pmatrix} 1 & 0 \\ 0 & -1 \end{pmatrix}$, and $\sigma_0$ is the identity matrix. Dispersions in $k_y$ direction are $\varepsilon_1 = (v + c_y)k_y$ and $\varepsilon_2 = (v - c_y)k_y$, whose crossing forms Dirac cone with Fermi velocities $v_{F1} = v + c_y$ and $v_{F2} = v - c_y$, respectively. The nodal point of the cone is known as Dirac point, whose energy is marked as $\varepsilon_D$. $v_{F1}$ and $v_{F2}$ can be tuned by changing $v$ and $c_y$, leading to three types of cones with distinct band dispersions and Fermi surfaces. For an upright type-I Dirac cone, $|c_y| \gg |v|$, $v_{F1} = -v_{F2}$. When the upright cone tilts, $|c_y|$ and $|v|$ decreases, and we elaborate here on a clockwise tilt having $c_y > 0$ and $v < 0$ [see Fig. 2(a)]. The cone remains as type-I when $c_y$ decreases from $+\infty$ to $-v$ ($c_y > -v$). At the critical point ($c_y = -v$), the cone has a flat band of $\varepsilon_1$ ($v_{F1} = 0$, $v_{F2} < 0$), which is dubbed type-III with a line-like Fermi surface [24, 25]. Beyond the critical point, an overtilted cone is named type-II having $c_y < -v$, whose two Fermi velocities have the same sign ($v_{F1}, v_{F2} < 0$) and Fermi surface encompasses both electron and hole pockets.

We next transform the Dirac fermions in crystals to the particles in gravity field. In Einstein's notation, Eq. (1) can be rewritten as $H(\boldsymbol{k}) = e_j^i \sigma^j k_i + e_0^i \sigma^0 k_i$. $i, j = 1, 2$ (or $x, y$) and the matrix $e_j^i$ and vector $e_0^i$ are equivalent to components of a tetrad field $e_\alpha^\mu$ ($\alpha, \mu = 0, 1, 2$ in $2 + 1$ dimensions) in general relativity. Then, an effective relativistic covariant metric $g_{\mu\nu} = (\eta^{\alpha\beta} e_\alpha^\mu e_\beta^\nu)^{-1}$ with $\eta^{\alpha\beta} = \text{diag}(-1,1,1)$ governs Dirac fermions. The corresponding line element ($ds^2 = g_{\mu\nu} dx^\mu dx^\nu$) is

$$ds^2 = -\left(1-\frac{v^2}{c_y^2}\right)dt^2 + \frac{1}{c_x^2}dx^2 + \frac{1}{c_y^2}dy^2 - \frac{2v}{c_y^2}dtdy, \tag{2}$$

which shows the behavior of Dirac quasiparticles in an effective 2 + 1 dimensional ($t$, $x$, $y$) spacetime. For $v = 0$, $ds^2 = -c_y^2 dt^2 + dy^2$ in ($t$, $y$) spacetime, representing a flat spacetime where electronic wave propagates at the velocity of $c_y$ along $\pm y$ directions. Hence, $c_y$ corresponds to light velocity ($c$) in the gravitational spacetime, so that Dirac wave propagating in a crystal field is formally equivalent to light propagating in a gravity field. In contrast to a constant $c$, however, the "light velocity" ($c_y$) in a crystal can be changed by interactions.

Effective spacetime for a given type of Dirac fermion can be designed by the relative magnitude of $c_y$ and $v$. The Dirac cone manifests in its "spacetime" as an artificial light cone ($ds^2 = 0$) with $t_1 = \frac{y}{v+c_y}$ and $t_2 = \frac{y}{v-c_y}$, as shown in Fig. 2(b). For type-I, $t_1' = 1/(v + c_y) > 0$ and $t_2' = 1/(v - c_y) < 0$ having the opposite sign, quasiparticles propagate along both $+y$ and $-y$ directions. For type-III, $t_1 \to \infty$, one branch of quasiparticles stays at a fixed location (the event horizon) permanently while the other propagates along $-y$ direction ($t_2' < 0$). For type-II, both $t_1'$ and $t_2'$ are negative, all quasiparticles propagate along $-y$ direction resembling the unidirectional behavior of particles inside a SBH.

To produce a desired gravity field for Dirac fermions, an appropriate spatial distribution of effective geometry is designed. Matching the effective and Schwarzschild metrics [31], the effective potential energy of quasiparticles is $\Phi(y) = -\frac{1}{2}\frac{v^2(y)}{c_y^2(y)}$. $\Phi(y)$ is inversely proportional to the distance: $\Phi(y) = -\frac{1}{2}\frac{y_h}{y}$, where $y_h$ is the location of event horizon. Consequently, the effective light velocity $c_y$ distributes along the $y$ direction as

$$c_y(y) = -v\sqrt{\frac{y}{y_h}}. \tag{3}$$

This guarantees that type-II fermions are inside the BH ($0 < y < y_h$), type-III at the event horizon ($y = y_h$) and type-I outside the BH ($y > y_h$) [Fig. 2(c)]. The motion of quasiparticles is described by kinematic equations of $k_{y1} = \frac{\varepsilon_1}{v(1-\sqrt{\frac{y}{y_h}})}$ and $k_{y2} = \frac{\varepsilon_2}{v(1+\sqrt{\frac{y}{y_h}})}$. As shown in Fig.

2(d), $k_{y1}$ and $k_{y2}$ represent two ingoing waves inside the BH, and one ingoing and one outgoing wave outside the BH. At the event horizon, there is such a high potential barrier that quasiparticles occupying on the $\varepsilon_1$ band are impossible to go through classically, but can tunnel through via quantum fluctuation to produce Hawking radiation.

Next, using quantum tunneling method [29, 32], we analyze the analogous Hawking radiation at the effective BH event horizon with the curved geometry of Eq. (3). As illustrated in Fig. 2(c), the $\varepsilon_1$ states above the Dirac point ($\varepsilon_1 > \varepsilon_D$) are occupied inside the BH, but empty outside. This leads to emission of the excited electrons from inside to outside the BH. Adopting Wentzel-Kramers-Brillouin approximation, the tunneling probability is $P = 1/(1+\exp(2S))$ with a classical action $S = \mathrm{Im} \int k_y(y)dy$. Assuming $k_y \approx \varepsilon \left/ \left( \left.\dfrac{dc_y}{dy}\right|_{y_h} \cdot (y-y_h) \right) \right.$ for the $k_{y1}$ branch around the event horizon, $P$ produces a spectrum with the energy intensity $I(\varepsilon) = n^e_{\mathrm{rad}}(\varepsilon)\cdot\varepsilon$. The number of radiated electrons follows

$$n^e_{\mathrm{rad}}(\varepsilon) \propto \frac{\varepsilon}{\exp\left(\dfrac{\varepsilon}{k_B T_H}\right)+1}, \qquad (4)$$

where $k_B$ is the Boltzmann constant, and $T_H = \dfrac{1}{2\pi k_B}\left|\dfrac{d}{dy}(c_y - v)\right|_{y_h}$ is the Hawking temperature. The energy spectrum $I(\varepsilon) = n^e_{\mathrm{rad}}(\varepsilon)\cdot\varepsilon$ of radiated massless Dirac electrons conforms to the thermal radiation of massless Dirac particles from a 2D gravitational BH (see Supplemental Material [33]). The "radiated" electrons and holes created by quantum fluctuation are entangled (lower panel in Fig. 1). It will also be very interesting to study the correlation effects between a Hawking pair (electron-hole pair) using our proposed fermionic analog of BH, such as by detecting and analyzing the distributions of both electrons (in the type-II region) and holes (type-I region) simultaneously resulted from quantum tunneling.

The radiated Hawking spectrum $I(\varepsilon)$ can be measured from the local electron distribution $n(\varepsilon)$ in the region $y > y_h$, using scanning tunneling spectroscopy or angle resolved photoemission spectroscopy (ARPES). As shown in Fig. 2(e), the energy of $\varepsilon = 0$ is set as the Fermi level ($\varepsilon_F$), at which Dirac point and chemical potential locate ($\varepsilon_D = \varepsilon_F = 0$). The spectrum $n^e_{\mathrm{rad}}(\varepsilon)$ with $\varepsilon > \varepsilon_D$ has a peak at $\varepsilon_p = 0.3$ meV for $T_H = 3$ K, which is distinct from the Fermi-Dirac distribution; and $n^h_{rad}(\varepsilon)$ for radiated holes below $\varepsilon_D$ due to quantum tunneling is similar. The Hawking spectrum (in 2D) derived here is the same as that of massless fermions emitted from a 2D BH, providing a key signature for Hawking radiation. $T_H$ can be tuned by controlling the effective gravity field to facilitate the experimental detection of $n^e_{rad}(\varepsilon)$ and $n^h_{rad}(\varepsilon)$. The relation between $\varepsilon_p$ and $T_H$ follows $\frac{\varepsilon_p}{k_B T_H} = 1 + \exp\left(-\frac{\varepsilon_p}{k_B T_H}\right)$, leading to a linear dependence $\varepsilon_p = 1.28 \cdot k_B T_H$ [Fig. 2(f)]. An unprecedentedly high $T_H$ is achieved because interactions in crystals are orders-of-magnitude stronger than those in gravity field and Bose-Einstein condensate.

Next, we theoretically design a solid-state system to realize such a fermionic analogue of BH. By applying a gate with the vertical electric field $E_{ext}$ or compressive strain $\delta$ along armchair ($x$) direction, the direct band gap of 2D BP thin film decreases [25, 34-40], leading to an inversion $\Delta\varepsilon$ of valence ($\varepsilon_1$) and conduction ($\varepsilon_2$) bands. Symmetry-protected type-I Dirac cone emerges, which was confirmed by ARPES [37, 38]. For a bilayer BP, this Dirac state is shown in Fig. 3(c), where $\Delta\varepsilon = 14$ meV at Γ point is induced by $E_{ext} = 0.18$ V/Å or $\delta = 7.6$ % (see Fig. S1 for details [33]).

To form the three types of Dirac states by laser-driving, we study coherent interactions between the bilayer BP and a linearly polarized laser (LPL) with a time-dependent vector potential $A(t) = A_0\sin(\omega t, 0, 0)$ [Fig. 3(a)]. The photon energy of the time-periodic and space-homogeneous LPL is chosen as $\hbar\omega = 0.03$ eV, which is larger than $\Delta\varepsilon = 14$ meV to avoid crossing nearby Dirac point between the original ($n = 0$) and photon-dressed ($n \neq 0$) bands. When the LPL is applied, the type-I Dirac cone tilts due to the hybridization between $n = 0$ and $n \neq 0$ bands. At the critical laser amplitude $A_0 = 16$ V/$c$ (corresponding to 0.24 mV/Å or $7.65 \times 10^5$ W/cm$^2$, here $c$ is light velocity), $\varepsilon_1$ band along ΓY path becomes flat, forming type-III Dirac cone [Fig. 3(d)]. As the laser amplitude increases to $A_0 = 20$ V/$c$, the Dirac cone tilts further to become type-II [Fig. 3(e)]. Then the slope of $\varepsilon_1$ and $\varepsilon_2$ dispersions has the same sign, and the states of the $\varepsilon_1$ band above Fermi level ($\varepsilon_1 > 0$ eV) becomes occupied. Consequently, type-I, II and -III Dirac fermions are created in a single material of 2D BP by a varying laser intensity. In addition, laser frequency plays also a crucial role in determining the type of cones, which is shown in the phase diagram of Fig. 3(b).

To mimic the spacetime geometry of Eq. (2), we map the photoinduced Dirac states from *ab initio* calculations to model parameters in Eq. (1). As shown in Fig. 3(f), the parameter $v = -0.1$ eV·Å is constant, while $c_y$ decreases gradually with increasing $A_0$, showing the transition from type-I ($c_y > -v$) to type-II ($c_y < -v$). In the regime of strong (weak) laser intensity, $c_y$ and $A_0$ exhibit a linear (nonlinear) relation, which is a typical characteristics of optical Stark effect (see details in Fig. S3 [33]). Neglecting the small nonlinearity, one can fit $c_y$ and $A_0$ as

$$c_y[\text{eV·Å}] = 0.314 - 0.013 A_0[\text{V}/c], \tag{5}$$

which is the base to explore Hawking radiation in the laser-driven bilayer BP thin film.

Substituting $c_y$ in Eq. (3) with Eq. (5), the amplitude of space-inhomogeneous laser field is $A_0(y)[V/c] = -8.154\sqrt{\dfrac{y[\text{Å}]}{y_h[\text{Å}]}} + 24.1538$, leading to $T_H = \dfrac{-v}{4\pi k_B y_h}$. Similar to $T_H \propto \dfrac{c}{r_h}$ for a gravitational BH, the $T_H$ of the proposed artificial BH is inversely proportional to the size ($y_h$) of a BH. In the cases of gravitational BH, $r_h$ is determined by the mass of BH, $r_h = 2GM$, which is large producing extremely low $T_H$. Here we could intentionally control the focus area of light with suitable intensity to achieve a small $r_h$ and thus a high $T_H$. To achieve $T_H = 3$ K, the BH size is set at $y_h = 30.2$ Å, which requires the laser field to decrease from 0.37 to 0 mV/Å in a range of 260 Å along the zigzag ($y$) direction of BP [Fig. 4(a)]. This gradient of laser intensity can be readily realized in experiments [46-48]. This temperature is realizable based on typical conditions of the laser experiment, as well as the validity of lattice periodicity as required by Bloch band theory. The radiation can be detected by measuring electron distribution $n(\varepsilon) = n_0(\varepsilon) + n^e_{rad}(\varepsilon) - n^h_{rad}(\varepsilon)$ in the region $y > y_h$, where $n^e_{rad}(\varepsilon)$ and $n^h_{rad}(\varepsilon)$ are the radiated spectrum of electrons and holes in Fig. 2(e), respectively, and $n_0(\varepsilon)$ is the electron distribution before radiation. As shown in Fig. 4(b), the peak position $\varepsilon_p$ of the calculated distribution $n(\varepsilon)$ is 0.3 meV above the Fermi level. Decreasing the BH size, $\varepsilon_p$ increases in a wide range of $T_H$ [Fig. 4(c)]. Also, we note that there are a pair of Dirac cones along –YY path resembling black and white holes respectively, but their coexistence does not influence the Hawking radiation (see details in Fig. S4 [33]).

In conclusion, type-I, -II and -III Dirac fermions are predicted to form in 2D BP thin film under LPL, resulting in a fermionic analogue of BH Hawking radiation. Hawking temperature is theoretically estimated to reach 3 K, with the emitted electrons exhibiting a spectrum peaked at 0.3 meV above the Fermi level, which should be experimentally observable. Our finding opens a path to engineering the table-top fermionic condensed-matter platform for simulating exotic phenomena in astrophysics and general relativity.

We thank financial support from the National Key Research and Development Program of China (Grants No. 2016YFA0300902 and No. 2016YFA0202300), National Basic Research Program of China (Grant No. 2015CB921001), NSF of China (Grants No. 11774396 and No. 11474328), and "Strategic Priority Research Program (B)" of CAS (Grants No. XDB30000000 and No. XDB07030100). H. H. and F. L. were supported by U.S. DOE-BES (Grant No. DE-FG02-04ER46148).

* jtsun@iphy.ac.cn
† fliu@eng.utah.edu
‡ smeng@iphy.ac.cn

# Figures

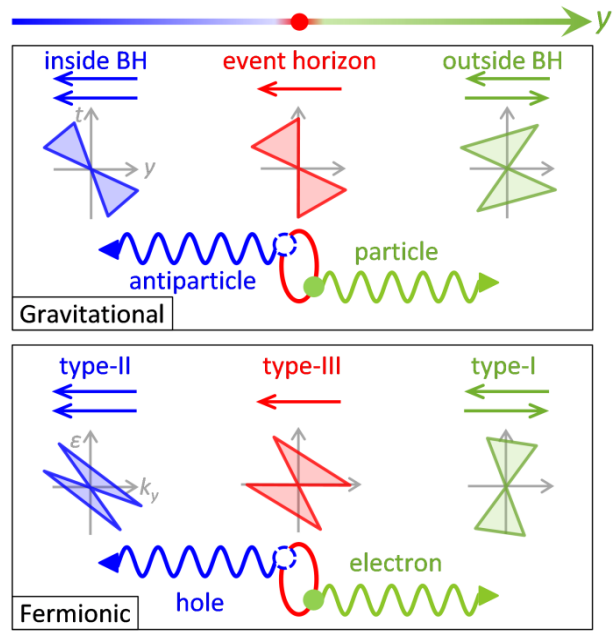

FIG. 1. Schematic illustration of gravitational Hawking radiation and fermionic analog. Upper panel: Light cones in real space indicate that particles cannot escape from gravitational BH classically (arrowed straight line), but particles and antiparticles can emit from event horizon due to quantum fluctuation (arrowed wavy line). Lower panel: Dirac cones in momentum space show that electrons cannot escape from the type-II region (arrowed straight line) of the artificial BH in crystals, quantum tunneling enables emission of electrons and holes from the event horizon (type-III region) to type-I and -II regions respectively (arrowed wavy line).

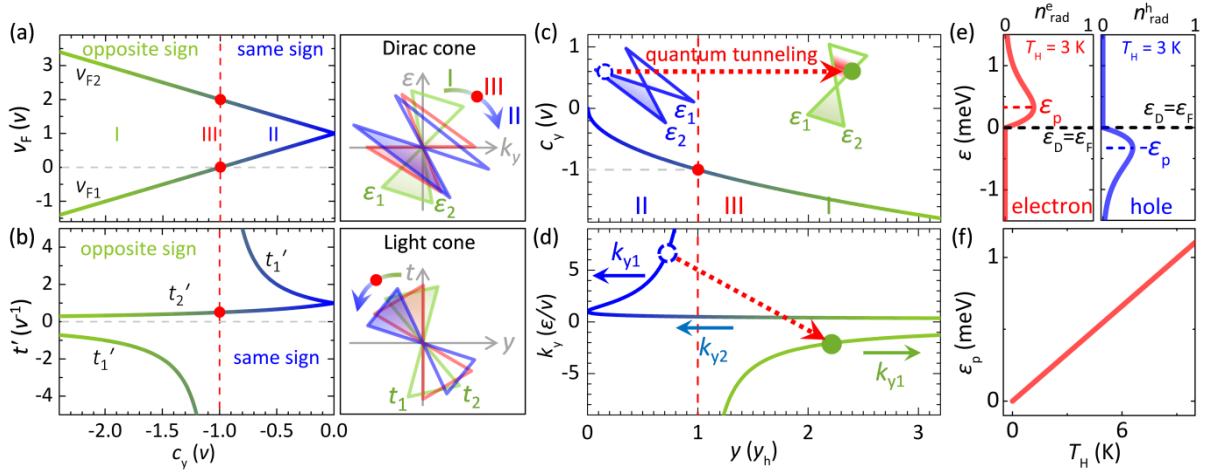

FIG. 2. Schematic illustration of fermionic analogue of a BH and consequent Hawking radiation. (a) Fermi velocity $v_F$ of Dirac cone varies with $c_y$, resulting in the transition from type-I (green), type-III (red) to type-II (blue) Dirac cone successively. (b) Three types of artificial light cone (green, red, and blue) formed via counterclockwise rotation, corresponding to type- I, -III, and -II Dirac fermions respectively. (c) The distribution of $c_y$ of type-I, -III, and -II Dirac fermions appears successively along $+y$ direction, producing the same "gravity field" felt by electrons as that of particles in a SBH. (d) Kinematic equation shows a high potential barrier at $y_h$ to prevent electrons from escaping from type-II to type-I region, but this process takes place by quantum tunneling indicated by the arrowed red dashed line in (c) and (d). (e) The spectrum of electrons (left) and holes (right) produced by Hawking radiation with $T_H = 3$ K. (f) The relation between the peak position $\varepsilon_p$ in (e) and $T_H$.

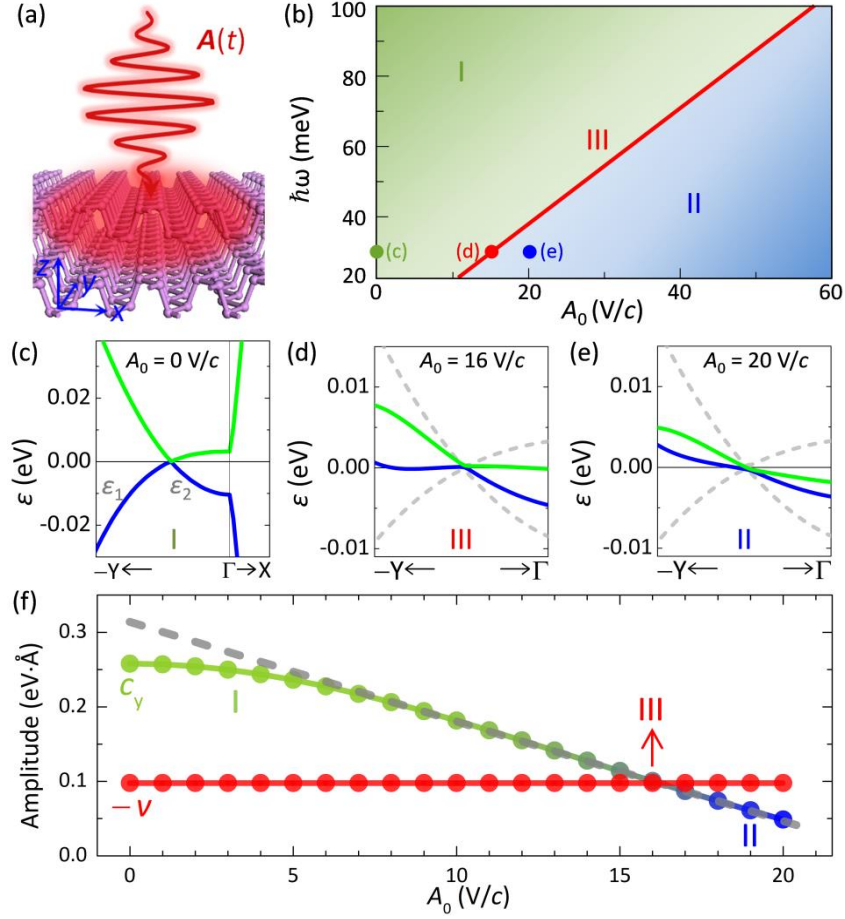

FIG. 3. First-principles calculated photoinduced type-I, -II and -III Dirac fermions in bilayer BP. (a) Laser field $A(t) = A_0(\sin(\omega t),0,0)$ is polarized along armchair ($x$) direction. (b) Dirac states induced by laser with different amplitude $A_0$ and frequency $\omega$. (c) Band structure of bilayer BP under vertical static electric field of 0.17 V/Å (or 7.6% compressive strain). Floquet-Bloch band structure of bilayer BP driven by laser with $A_0 = 16$ V/$c$ (d) and $A_0 = 20$ V/$c$ (e), and photon energy $\hbar\omega = 0.03$ eV. (f) With increasing laser intensity, type-I Dirac fermion is transitioned to be type- III, and -II successively. The gray dashed line shows linear fitting of $c_y$ vs. $A_0$.

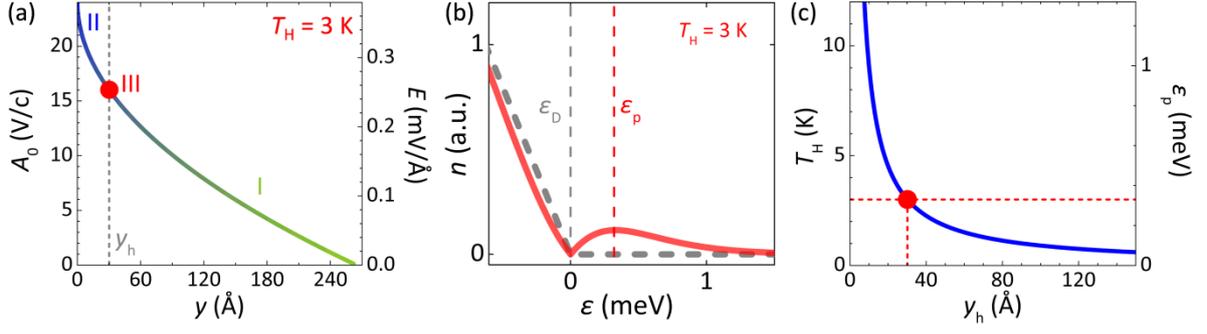

FIG. 4. The theoretically designed Hawking radiation in a bilayer BP thin film. (a) Required laser intensity distribution along zigzag ($y$) direction to realize a SBH with $T_H = 3$ K. (b) The electron distribution $n(\varepsilon)$ at $y > y_h$ after Hawking radiation with $T_H = 3$ K. The $n_0(\varepsilon)$ before Hawking radiation is shown by the gray-dashed line. (c) The calculated dependence of $T_H$ on the location of event horizon. The red dot corresponds to $T_H = 3$ K and $\varepsilon_p = 0.33$ meV.